\documentclass[twocolumn,secnumarabic,amssymb,nobibnotes,iop,showpacs,nofootinbib,noeprint]{revtex4-1}
\usepackage[english]{babel}
\usepackage{graphicx}

\usepackage{amsmath}
\usepackage{amsfonts}
\usepackage[usenames,dvipsnames]{color}
\usepackage{hyperref}
\usepackage{rotating}
\usepackage{slashed}
\usepackage{transparent}

\DeclareMathAlphabet{\mathbb}{U}{bbold}{m}{n}

\begin{document}
\title{Anisotropy and memory during cage breaking events close to a wall}
\author{Matthias Kohl$^1$} 
\author{Andreas H\"{a}rtel$^2$} 
\email{AnHaerte@uni-mainz.de}
\author{Michael Schmiedeberg$^{3}$} %
\email{michael.schmiedeberg@fau.de}
\affiliation{$^1$ Institute for Theoretical Physics II: Soft Matter, 
	Heinrich Heine University D\"usseldorf, Universit\"atsstr. 1, D-40225 D\"usseldorf, Germany \\
$^2$ Institute of Physics, Johannes Gutenberg-University Mainz, Staudinger Weg 9, 55128 D-Mainz, Germany\\
$^3$ Institute for Theoretical Physics I, Friedrich-Alexander-University Erlangen-N\"urnberg, Staudtstr. 7, 
D-91058 Erlangen, Germany}

\begin{abstract} 
The slow dynamics in a glassy hard-sphere system is dominated by cage breaking events, 
i.e., rearrangements where a particle escapes from the cage formed by its neighboring particles. 
We study such events for an overdamped colloidal system 
by the means of Brownian dynamics simulations. While it is difficult to relate 
cage breaking events to structural mean field results in bulk, we show that the microscopic dynamics 
of particles close to a wall can be related to the anisotropic two-particle density. 
In particular, we study cage-breaking trajectories, mean forces on a tracked particle, and the 
impact of the history of trajectories. Based on our simulation results, we further construct two different 
one-particle random-walk models -- one without and one with memory incorporated -- 
and find the local anisotropy and the history-dependence of particles 
as crucial ingredients to describe the escape from a cage. 
Finally, our detailed study of a rearrangement event close to a wall not only reveals the 
memory effect of cages, but leads to a deeper insight into the fundamental mechanisms of glassy dynamics. 
\end{abstract}

\maketitle

{\it This is the Accepted Manuscript version of an article accepted for 
publication in Journal of Physics: 
Condensed Matter. IOP Publishing Ltd is not responsible for any errors 
or omissions in this version of the manuscript or any 
version derived from it. The Version of Record is available online at 
http://dx.doi.org/10.1088/0953-8984/28/50/505001. }

\section{Introduction}

Hard spheres with solely steric repulsion are an important model system in the field of soft 
condensed matter as they exhibit not only a fluid and a crystalline phase \cite{hoover1968}, 
but there occurs a dramatic slowdown of dynamics when also the density of the particles is 
increased \cite{woodstock1976,Pusey1986}. The relation of this slow dynamics within a glassy system 
to structural properties is an important topic of ongoing (theoretical) research 
\cite{Kirkpatrick1985,Gotze1992,Santen2000,Schweizer2004,Schweizer2005,Elmatad2009,Granato2011,Mirigian2013,PhysRevE.92.052304} 
and aims on a deeper understanding of the major mechanisms that dominate glassy dynamics. 

\begin{figure*}
\centering
\includegraphics[width=13.0cm]{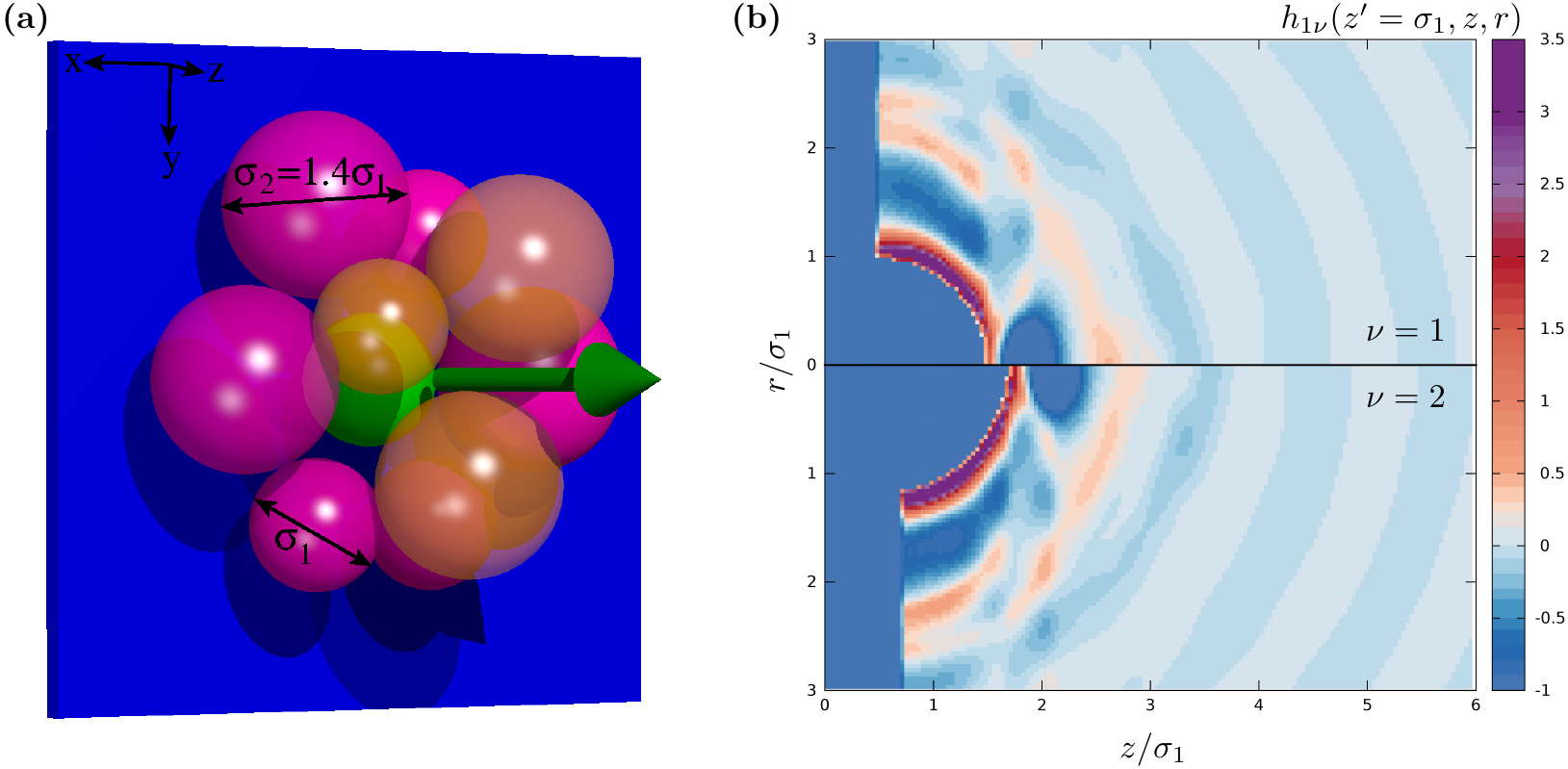}
\caption{\label{fig:intro}(a) Schematic sketch of a particle (shown in green) with diameter $\sigma_1$ 
close to a wall. The particle is trapped in a cage formed by its neighbors with diameters $\sigma_1$ or 
$\sigma_2=1.4\sigma_1$. The $z$ axis is aligned perpendicular to the wall plane, which is located at 
$z=0$. Note that the coordinate system in the figure only denotes the directions of the axis. The origin 
is located at the wall at the projected position of the trapped particle, i.e., the latter is located at 
the radial coordinate $r=\sqrt{x^2+y^2}=0$. 
(b) Two-particle pair correlation function as studied in previous work \cite{haertel2015} by DFT 
and BD simulations. Here, we show data from DFT for a total packing fraction $\phi=0.56$, 
where a small particle is located at the wall and its neighbors form a cage. 
The upper panel shows the correlations between the trapped particle and the other small particles, 
the lower panel those between the trapped particle and the large particles. More details and further 
examples on the structure of such a system can be found in \cite{haertel2015}. }
\end{figure*}

Many theoretical approaches to describe glassy dynamics are based on cage breaking events, e.g., the widely-used mode coupling theory \cite{Gotze1992,Gotze2008} or models that consider activated hopping of particles 
out of a cage 
\cite{Schweizer2004,Schweizer2005,Mirigian2013,PhysRevE.92.052304}.
Such mean-field theories are usually 
based on an isotropic average cage around the considered particle such that effects arising 
from the anisotropic structure of a cage are ignored. To avoid this drawback, we consider a cage close to a flat wall such that isotropy is broken. An example of a particle that is trapped by a cage of neighboring particles close to a wall is sketched in Fig.~\ref{fig:intro}(a). Since isotropy is broken, we have the advantage to know the 
preferred escape route from the cage (indicated by an arrow in Fig.~\ref{fig:intro}(a)). Our goal is to identify the essential ingredients that are 
necessary to develop an one-particle random-walk model that is able to describe the escape dynamics. 
Furthermore, our analyzes of the anisotropic cage-breaking event also opens another view on 
cage-breaking events in bulk, where cages are only isotropic in a mean field description due to 
averaging over different directions. Such an averaging might lead to mistakes when the dynamics 
along escape routes should be predicted. 

By using (classical) density functional theory (DFT), the anisotropic two-particle 
correlation functions and, therefore, the structure of a cage that traps a particle close to 
a wall can be determined theoretically \cite{dietrich1996,haertel2015} 
(the full gamut of classical DFT is covered in \cite{evans_jpcm28_2016}). As we will show later, 
structural results will be important to understand the anisotropic cage breaking dynamics.
For the system that we consider here, we have resolved the anisotropic structures in a previous 
work \cite{haertel2015}, where we studied a 
glass-forming binary hard-sphere mixture close to a wall by applying the White Bear mark II framework of 
fundamental measure theory \cite{hansen-goos_jpcm18_2006}, 
a quantitative benchmark DFT for hard spheres \cite{haertel_prl108_2012,oettel_pre86_2012}. 
We found very good agreement between DFT 
and Brownian dynamics (BD) computer simulations and could observe cage-forming structure. 
A typical result is shown in Fig.~\ref{fig:intro}(b), where the total 
correlation function (the pair distribution function follows by adding a constant) is 
shown for a small particle in contact with the wall. Both the correlations with neighbouring 
small (upper panel) and large (lower panel) particles have preferred positions for finding 
a neighbouring particle. This anisotropic structure can not be found from solely one-particle 
correlations, which are often studied in order to understand surface free energies and the 
adsorption of particles to walls 
\cite{malijevsky_pre75_2007,haertel_prl108_2012,davidchack_cmp19_2016}.

In Sec.~\ref{sec:model} we introduce our model system and explain details of the Brownian 
dynamics simulations. The dynamical self-correlation functions of a cage-braking event are determined 
and discussed in Sec.~\ref{sec:van-hove}. In Sec.~\ref{sec:properties}, we explore the properties 
of a cage, namely the force that acts on a caged particle due to its anisiotropically distributed 
neighbours, the collision 
frequency when the trapped particle inside tries to escape, and the history-dependence of the 
cage-breaking event. These properties are used as ingredients to develop one-particle random-walk 
models as explained in Sec.~\ref{sec:rw}. Finally, we conclude in Sec. \ref{sec:conclusions}.

\section{Model system}
\label{sec:model}

In this work we study a binary mixture of purely repulsive soft spheres 
in a solvent and in the vicinity of a wall by means of BD simulations, i.e., the 
solvent is captured by considering random kicks and in contrast to 
Newtonian dynamics no inertia effects occur. 
Half of the particles possess the effective diameter $\sigma_1$ and 
the other half has a larger diameter $\sigma_2=1.4\sigma_1$. Usually we consider $32000$ particles in 
a three dimensional cubic box with periodic boundary conditions in two directions (the $x$- and $y$-direction in our choice of the coordinate system) and closed by two walls normal to third direction (the $z$-direction).
We typically combine the number densities $\bar{\rho}_{\nu}$ of each species $\nu$ in a total 
packing fraction $\phi=\sum_{\nu}\tfrac{\pi}{6}\sigma_{\nu}^3\bar{\rho}_{\nu}$. 
The spheres, as sketched in Fig.~\ref{fig:intro}(a), interact according to the potential 
\begin{align}
	u_{\nu\nu'}(\Delta) &= \left\{ \begin{array}{lcl} 
	\frac{\varepsilon}{2}\left( 1 - \frac{\Delta}{\sigma_{\nu\nu'}}\right)^2 & \quad & \Delta\leq \sigma_{\nu\nu'} \\
	0 & & \text{otherwise} , 
	\end{array} \right. 
 \label{eq:pair_potential} 
\end{align}
where $\Delta$ is the separation of two interacting spheres, 
$\sigma_{\nu\nu'}=(\sigma_\nu + \sigma_{\nu'})/2$ involves the effective particle diameters 
$\sigma_{\nu}$, and $\varepsilon$ denotes the strength of the interaction. 
The particle-wall interaction is given by a similar harmonic potential with the same strength 
factor $\varepsilon$. In the limit of small temperatures $T$ and large interaction strengths, 
i.e. for $\varepsilon/k_\text{B}T\gg 1$ with Boltzmann's constant $k_{\rm B}$, 
the structure and dynamics of the system correspond to the structure and dynamics of a hard sphere 
system \cite{rowlinson64,barker67,Andersen1971,Xu2009,Schmiedeberg2011,Haxton2011,Lopez-Flores2013,haertel2015}. 
We have chosen $\varepsilon$ such that for a given temperature the average overlap between 
two particles or between a particle and the wall does not exceed $5\%$ of the diameter 
$\sigma_1$ (see also \cite{haertel2015}). 

The dynamics of our BD simulations is given by the overdamped Langevin equation 
\begin{align}
\gamma_\nu \dot{\vec r}_{\nu,i}(t) &= {\vec f}_{\nu,i}\big(\left\{\vec r_{\nu',1},\ldots, 
\vec r_{\nu',N_{\nu'}}\right\}_{\nu'=1,2,...}\big) + \vec \xi_{\nu,i}(t)\, . \label{eq:langevin}
\end{align}
The friction constants $\gamma_\nu$ are proportional to the diameter $\sigma_{\nu}$ of the spheres. 
The pair interaction forces between the particles and between the particles and the walls are taken 
into account in ${\vec f}_{\nu,i}$. The thermal kicks due to a given temperature $T$ are included 
by random forces $\vec \xi_{\nu,i} (t)$ that are chosen from a Gaussian distribution with zero mean 
value and second moments given by 
$\big\langle \vec \xi_{\nu,i}(t) \vec{\xi}_{\nu',i'}^\text{T}(t') \big\rangle 
= 2 \gamma_\nu k_{\rm B} T \delta_{\nu\nu'}\delta_{ii'}\delta(t-t')\Bbb{1}$, 
where $\vec{\xi}_{\nu'}^\text{T}$ is the transpose of $\vec{\xi}_{\nu'}$ and 
$\Bbb{1}$ is the three-dimensional unit matrix. We employ $\tau_B=\sigma_1^2/(3\pi D_1)$ 
as a suitable Brownian time throughout the article, where $D_1$ is the coefficient of free 
diffusion of the small spheres. More details about the simulations and an analysis of the hard sphere limit are given in a previous 
work \cite{haertel2015}, where we studied the structure of a binary hard-sphere system in the vicinity 
of a hard wall by simulations and DFT calculations.

\section{Dynamical correlations of anisotropic cage-breaking events}
\label{sec:van-hove}

In dense systems each particle is surrounded by a cage of neighboring particles 
(see for instance Fig.~\ref{fig:intro}). In bulk such a cage can possess any orientation and, 
for this reason, a mean field description usually is isotropic. Instead, close to walls 
a preferred orientation of the cage exists. It occurs more often than other configurations. 
Consequently, preferred escape routes exist for cage-breaking events, too. 
In this section we study dynamical time-dependent van Hove self-correlation functions obtained from 
simulations in order to explore the anisotropy of cages and to identify preferred escape routes. 

\begin{figure*}
\centering
\includegraphics[width=0.95\linewidth]{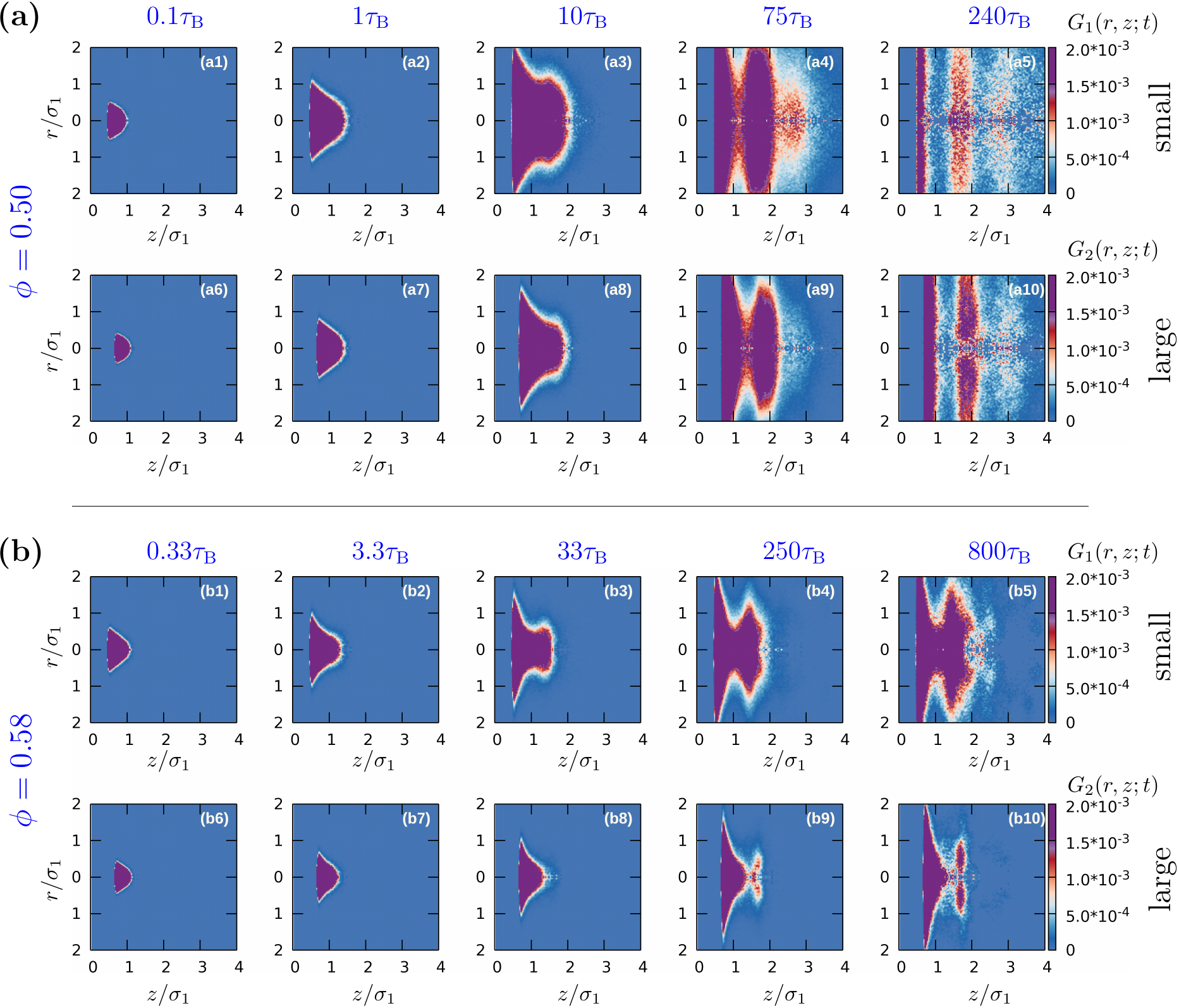}
\caption{\label{fig:vanHoveCompilation}Self parts of the van Hove correlation function 
$G_\nu(r,z;t)$ for (a) $\phi=0.50$ and (b) $\phi=0.58$. The cylindrical coordinates $z$ and $r$ represent 
the distance from the (left) wall located at $z=0$ and the distance from the $z$-axis, at which the traced 
particle is located initially at $t=0$. Plots from the left to the right show the functions 
$G_\nu(r,z;t)$ after different time (in units of the Brownian time $\tau_{\textnormal{B}}$), 
as indicated at the top of each column in panels (a) and (b). The upper row of each panel 
({\footnotesize\textsf{a1-a5}} and {\footnotesize\textsf{b1-b5}}) 
represents data for small particles ($\nu=1$), whereas the lower row of each panel 
({\footnotesize\textsf{a6-a10}} and {\footnotesize\textsf{b6-b10}}) represents data for the large 
particles ($\nu=2$). Note that the timespan in (a) is smaller than in (b). The color 
code is the same for all plots and given on the right side of each line.}
\end{figure*}

In Fig. \ref{fig:vanHoveCompilation}, we plot simulation results of the van Hove self-correlation 
function $G_\nu(r,z;t)$ \cite{hansen_book_2013}. 
The function represents the probability density for a particle of species $\nu$ 
to be at the position $(r,z)$ after a time $t$, if it was tracked in the origin $(0,0)$ at time $0$. 
Due to the symmetry of the system under investigation, we represent the position of a particle in 
cylindrical coordinates $(r,z)$, i.e. the distance $z$ from the wall and the distance $r$ from the $z$-axis 
(cf. Fig.~\ref{fig:intro}). The van Hove function represents the information about the diffusion pathways 
or preferred particle trajectories. Figure~\ref{fig:vanHoveCompilation}(a) displays $G_1(r,z;t)$ for small 
particles (first row) and $G_2(r,z;t)$ for large particles (second row) at a volume fraction $\phi=0.50$ 
and for different evolution times, as indicated. The caging of particles is visible in the slowed down 
diffusive dynamics for shorter times (Fig.~\ref{fig:vanHoveCompilation}, {\footnotesize(\textsf{a1-a3})} 
and {\footnotesize(\textsf{a6-a8})}). Of course, diffusion is prohibited inside the wall, but the overall 
diffusion process in all remaining directions still looks rather homogeneous as long as the particle does not 
leave its cage.  
For a density close to the glass transition (see Fig.~\ref{fig:vanHoveCompilation}(b) for 
$\phi=0.58$), $G_\nu(r,z;t)$ displays an even slower diffusion process (note the different timescales). 
On the same time, the diffusion within the cage becomes more anisotropic, e.g., while in 
Fig.~\ref{fig:vanHoveCompilation}{\footnotesize(\textsf{a2})} no preferred direction of motion is visible, 
close to the glass transition as shown in Fig.~\ref{fig:vanHoveCompilation}{\footnotesize(\textsf{b2})} 
a particle within a cage more likely moves parallel to the wall and the motion in all other directions 
is suppressed. 
In all cases, the small particles diffuse faster as expected from the smaller friction constant 
$\gamma_{1}$. 

For sufficiently long times, we can identify different escape routes along which the particle can 
leave the cage. Besides a diffusive spread within the first layer in direction parallel to the wall, 
smaller particles tend to use a perpendicular path to swap into the second layer 
(Fig.~\ref{fig:vanHoveCompilation}, {\footnotesize(\textsf{b2-b3})}) before they spread in parallel 
directions again (Fig.~\ref{fig:vanHoveCompilation}, {\footnotesize(\textsf{b4-b5})}). 
For the large particles, however, the typical diffusion paths are different. First, when 
large particles approach the second layer, they spread much slower along the second layer 
(Fig.~\ref{fig:vanHoveCompilation}, {\footnotesize(\textsf{b6-b8})}) than the small particles do. 
Second, when large particles arrive at a distance of approximately $1.7\sigma_1$, their most 
likely trajectories branch away from the $z$-axis, while small particles also stay at the $z$-axis 
(Fig.~\ref{fig:vanHoveCompilation}, {\footnotesize(\textsf{b9-b10})} and {\footnotesize(\textsf{b4-b5})}). 
This can be explained by a preferred stacking of large particles above small particles. If a large 
particle was initially located next to a small one, it raises towards the second layer until it is 
possible to stack with its smaller neighbor. In order to do so, it leaves the actual perpendicular 
path. The last plot of the time series of the large particles' van Hove function 
(Fig.~\ref{fig:vanHoveCompilation}, {\footnotesize(\textsf{a10 and b10})}) suggests that first isolated 
peaks evolve before the probability density washes out in the second layer. 
These two mechanisms, namely the perpendicular diffusion of the small and the tilted diffusion of the 
large particles, account for local rearrangement processes. 
Such rearrangements also occur in the bulk case, but due to isotropy with no preferred direction. 
Due to our symmetry breaking wall, in our situation the 
probability densities $G_\nu(r,z;t)$ are anisotropic and bear pre-defined more probable diffusion 
paths due to the structure of the local neighborhood. 

\begin{figure*}
\centering
\includegraphics[width=16.0cm]{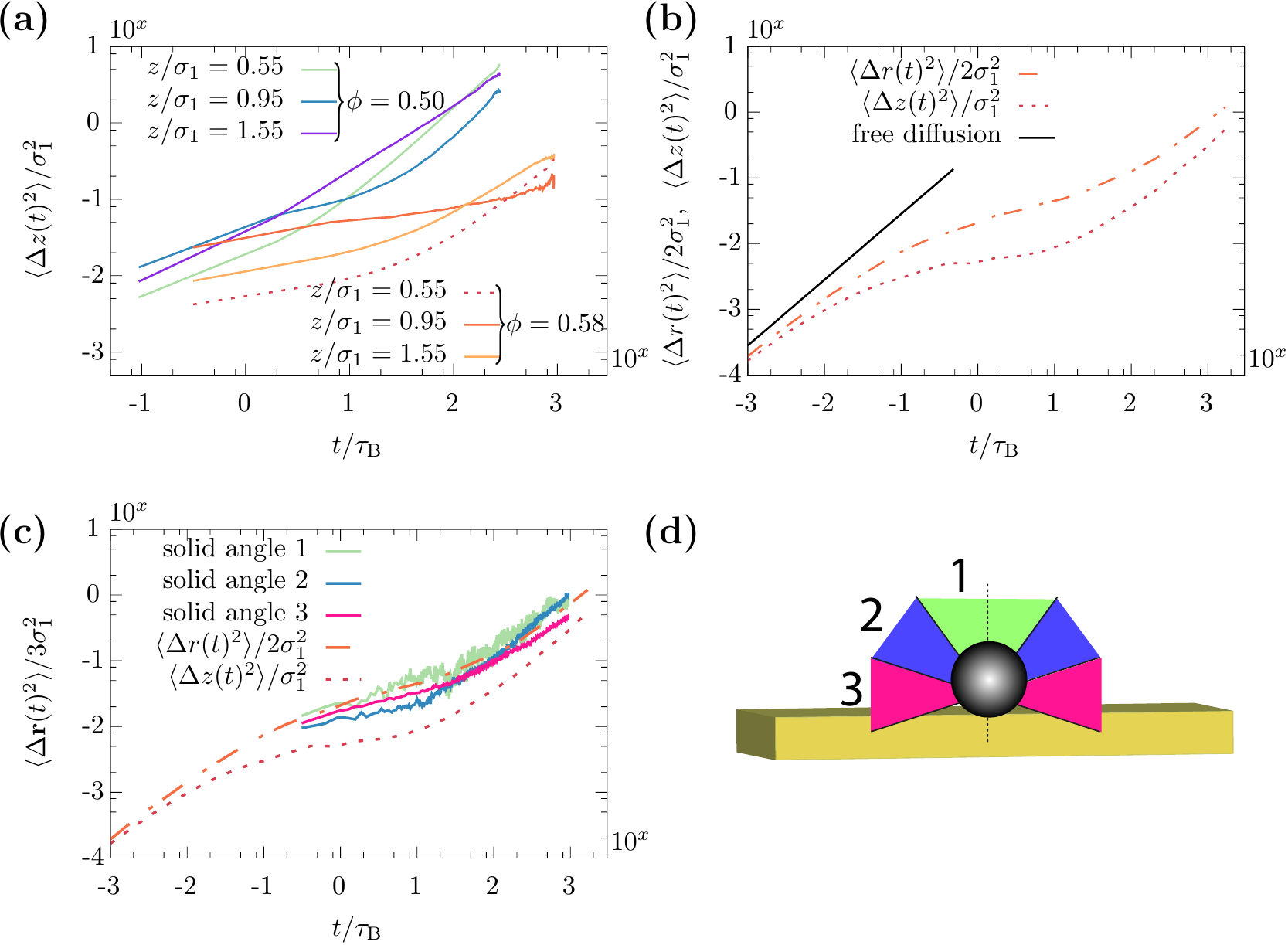}
\caption{\label{fig:MSDCompilation}Double logarithmic plots of the mean squared displacements (MSD) 
of small particles (species $\nu=1$) in a binary mixture. 
(a) The MSDs in $z$-direction for two packing fractions and 
three different starting positions of the respective trajectories. 
(b) MSD parallel ($\langle \Delta r^2 \rangle$) and perpendicular ($\langle \Delta z^2\rangle $) to 
the wall for a packing fraction $\phi=0.58$. For comparison, the theoretical curve of free diffusion 
is shown by a solid black line. (c) The dashed curves are the same as in (b), whereas the solid lines 
represent normalized radial MSD 
($\langle \Delta \vec r^2 \rangle_{[\vartheta_\text{I},\vartheta_\text{II}]}$), 
for which the averages have only been performed over a subset of particles that have moved into given 
directions within the solid angle ranges as sketched in (d) and labeled with (1,2,3). 
The intervals in the standard representation of the polar angle $\vartheta$ are 
(1)=$[0,\pi/5]$, (2)=$[\pi/5,2\pi/5]$, and (3)=$[2\pi/5,3\pi/5]$.} 
\end{figure*}

To quantify this effect we explore the consequences of the anisotropic escape routes for the directional 
dependence of the 
mean squared displacements (MSD). As we will show in the following, the diffusion process and the 
accompanied escape from the cage into a certain direction are similar to the diffusion and escape in all 
other spatial directions, but each event must be weighted with the probability for its occurrence. 
In Fig.~\ref{fig:MSDCompilation} we plot the normalized MSDs obtained from our BD 
simulations for small particles in different directions, i.e. in $z$-direction: 
\begin{align} 
\left\langle \Delta z(t)^2 \right\rangle = \frac{1}{N_1}\sum\limits_{i=1}^{N_1} \big(z_{i}(t)-z_{i}(0)\big)^2 , 
\end{align} 
in direction parallel to the wall (radial in the cylindrical coordinates): 
\begin{align} 
\left\langle \Delta r(t)^2\right\rangle = \left\langle \Delta x(t)^2\right\rangle 
+ \left\langle \Delta y(t)^2\right\rangle , 
\end{align} 
and in radial (spherical coordinates) direction in a certain solid angle range 
$[\vartheta_\text{I},\vartheta_\text{II}]$: 
\begin{align} 
\nonumber
&\left\langle \Delta \vec r(t)^2 \right\rangle_{[\vartheta_\text{I},\vartheta_\text{II}]} \\
&=\frac{1}{N_1^{'}}\sum\limits_{i=1}^{N_1} \theta\big(\vartheta_i(t) 
- \vartheta_\text{I}\big)\theta\big(\vartheta_\text{II} 
- \vartheta_i(t)\big)\big(\vec r_i(t) - \vec r_i(0)\big)^2 . \label{eq:MSD_pre_selection} 
\end{align} 
Here, we applied the Heaviside step function $\theta(\cdot)$ and the number of considered particles 
$N_1^{'}$, i.e., the number of non-zero addends, for particles that fulfill the criterion. 
Note that for simplicity we have omitted 
the index $\nu=1$ for small particles. Equation~(\ref{eq:MSD_pre_selection}) represents the diffusion 
process of selected particles that move only into the desired direction. Their final positions 
after a time $t$, denoted by the polar angles $\vartheta_i(t)$, are restricted to 
$\vartheta_\text{I} \leq \vartheta_i(t) \leq \vartheta_\text{II}$. 

The MSDs from Fig.~\ref{fig:MSDCompilation}(a) are calculated with respect to different starting 
positions, i.e., $z(t=0)/\sigma_1=0.55,0.95,1.55$, where in Fig.~\ref{fig:MSDCompilation}(b-d) the 
particles always start at wall contact. 
As it is clearly visible in Fig.~\ref{fig:MSDCompilation}(a), the effect of different starting positions 
has major impact on the shape of the vertical short-time MSD at intermediate or high packing fractions. 
If the tracked particle starts inside a layer, i.e., the maximum of the density profile 
(at $z_1/\sigma_1=0.55,\,1.55$), it immediately feels the impact of the confining cage at this position. 
This fact makes it to become trapped at short and intermediate times and only jump after larger waiting times. 
Furthermore, if a particle starts in the second layer, it is able to jump forward or backward. Therefore, 
especially at shorter times the diffusion is enhanced by a factor of approximately $2$ with respect to a 
particle that starts in front of the wall. 
However, the initial trapping in the cage is still visible. 
At larger times, backward jumps are restricted by the wall and the factor $2$ disappears. 
Contrarily, if the starting position is in between the first two layers ($z_1/\sigma_1=0.95$), 
the diffusion in $z$-direction is larger for short times and smaller for long times. This is due to the 
fact that the starting position for such a particle is less stable, because it is unfavorable to stay 
in between two layers. Successively, when the particle relocates along the $z$-axis, it inevitably 
reaches one of the neighboring layers, where it might be caught in a cage for a typical period. 
Since we aim to investigate the local rearrangement from one stable position to another, we will 
always consider particles that start at the wall in the following. 
 
In Fig.~\ref{fig:MSDCompilation}(b) we plot the MSD (averaged over all small particles) for a dense 
system with packing fraction $\phi=0.58$ for directions parallel and perpendicular to the wall in 
comparison with the free diffusion. The averaged curves suggest that the diffusion process in 
$z$-direction is stronger suppressed than parallel to the wall. 
This is because we consider the average over all particles in Fig.~\ref{fig:MSDCompilation}(b). 
However, if only particles within the pre-selected directions are employed, the different MSD look 
all rather akin as demonstrated in Fig.~\ref{fig:MSDCompilation}(c). All curves approximately 
correspond to the curve of the diffusion along the wall from Fig.~\ref{fig:MSDCompilation}(b). 
This intriguing finding leads to the conclusion that the diffusion behaves similar in all spatial 
directions, where at the same time diffusion along the wall is more probable (already seen 
from Fig.~\ref{fig:vanHoveCompilation}). Only because in the usual representation of the MSD 
the displacement is weighted by 
the probability of the accompanied sampling in a certain direction, a more pronounced plateau 
emerges when inspecting less probable directions, e.g., as in Fig.~\ref{fig:MSDCompilation}(b). 

One has to be careful with the interpretation of the above presented averages, 
since a particle could for example escape in a tilted direction due to the lack of a 
neighboring particle at that position, by chance. Such events would occur in the direction 
where the first and second layer in Fig.~\ref{fig:intro}(b) are connected by strongly correlated regions, 
i.e., along a path on which the minimal value of the total correlation function is as large as possible. 
Note that its direction is also captured by the solid angle interval 1 
in Fig.~\ref{fig:MSDCompilation}(d). 

Still, the above described perceptions motivate for the fact, that the escape in other directions 
can probably be treated in a similar way as the perpendicular escape, just with the difference 
that some directions are more probable than others. Concerning the importance for the bulk, 
the completely averaged diffusion process is probably the result of the weighted composition 
of the separate diffusion mechanisms in all possible directions. 

For this reason, we will focus on the cage escape in $z$-direction within our setup in the following 
sections. In particular, we will study the forces acting on a particle during its escape, relate the 
forces to collision frequencies in a hard-sphere system, and investigate the history dependence 
of escape paths.

\section{Properties of anisotropic cages and cage-breaking events}
\label{sec:properties}

In this section we analyze the forces acting on a sphere that is trapped in an anisotropic cage 
formed by its neighboring particles. Furthermore, we explore the history dependence of the dynamics 
during cage-breaking events. The properties studied in this section will be used as ingredients 
for the one-particle random-walk models proposed in Sec.~\ref{sec:rw}.

\subsection{Force distribution at finite temperatures}

\begin{figure*}
\centering
\includegraphics[width=13.0cm]{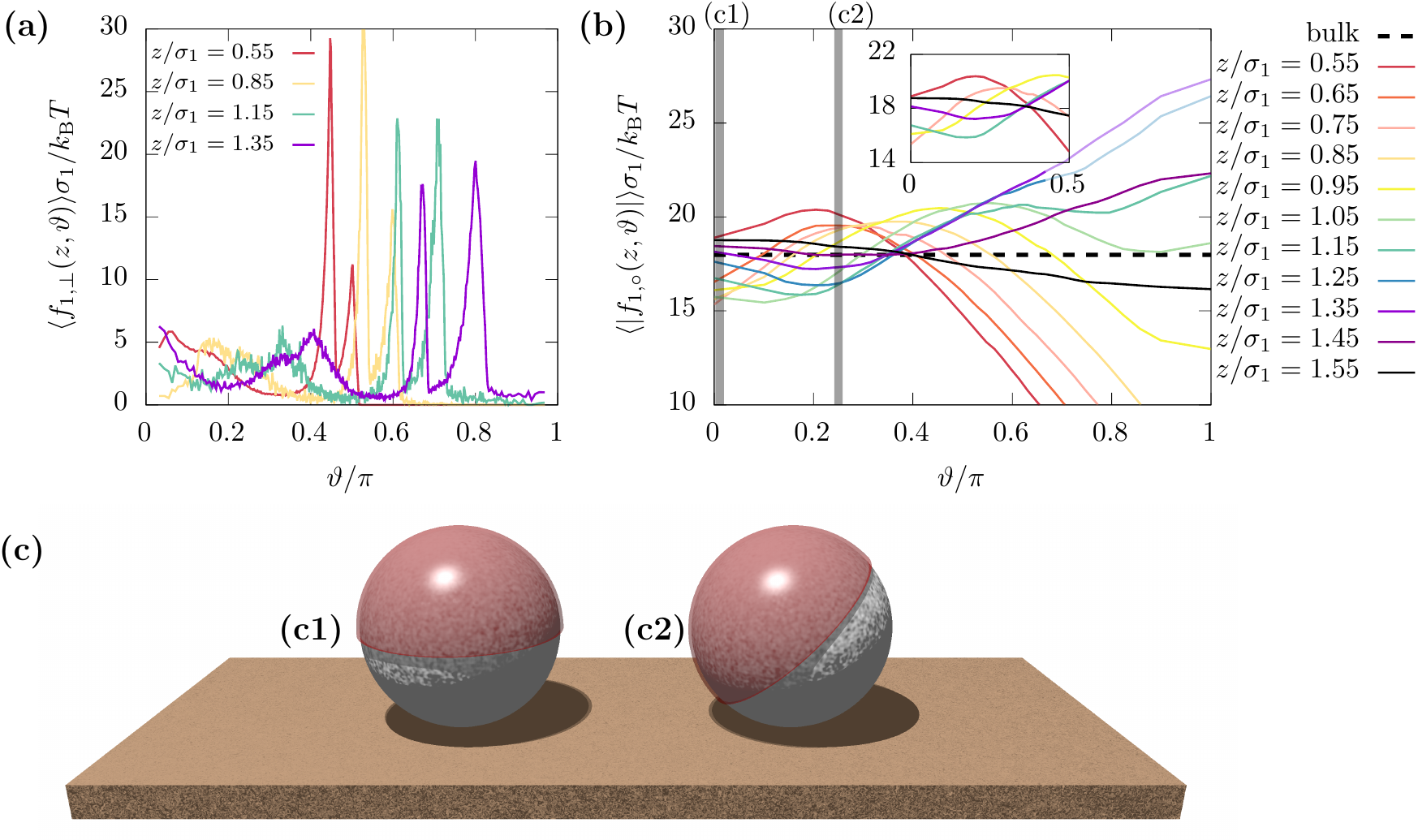}
\caption{\label{fig:ForceDistributionAndCaps}(a) Distribution of the mean normal forces 
$\langle f_{1,\perp}\rangle$ that act on a small particle from different directions in a 
binary mixture with overall packing fraction $\phi=0.58$ and for various distances $z$ from the wall. 
The angle $\vartheta$ denotes the direction of the force and is measured with respect to 
the $z$-axes, where $\vartheta=0$ corresponds to forces acting towards the wall. 
(b) Cross section forces $\langle f_{1,\circ}\rangle$ obtained by averaging over all force 
contributions that act along a direction denoted by $\vartheta$ on the half sphere pointing 
in that direction. This corresponds to the forces that the particle has to overcome in order 
to move in the corresponding direction. The direction-independent bulk value is given for 
comparison by a dashed black line; $z$ again denotes the distance from the wall. The inset 
highlights the behavior on hemispheres that point away from the wall. 
(c) Two examples for particles with marked hemispheres for ({\footnotesize\textsf{c1}}) 
$\vartheta=0$ and ({\footnotesize\textsf{c2}}) $\vartheta=\pi/4$. The corresponding positions 
are highlighted by vertical gray stripes in panel (b). 
}
\end{figure*}

While in the previous section we focused on the preferred trajectories which trapped particles 
use to escape from cages, we now look for mean forces on particles during their escape. 
In order to explore the properties of a cage, we use the BD simulations to calculate the 
mean forces $\langle f_{1,\perp}(z,\vartheta)\rangle$ that act on the surface of a small 
particle. These forces arise from overlaps with neighboring particles which are located in 
directions denoted by the polar angle $\vartheta$. 
Figure~\ref{fig:ForceDistributionAndCaps}(a) shows these forces averaged over a large number 
of configurations in a dense system with $\phi=0.58$ for various distances $z$ from the wall. 
The two very prominent peaks in all the distributions stem from the likely configurations where the 
small test particle is the neighbor of a small or a large sphere at the wall, respectively. 
As the tracked particle detaches from the wall, these peaks shift from the sides towards the 
bottom hemisphere ($\vartheta=\pi$) of its surface (if the wall is considered at the bottom of 
the box). However, from atop ($\vartheta=0$) the forces become smaller. 
This is due to the unfavored position of a neighboring particle on top of the test particle, because 
particles of the second layer from the wall are usually not positioned directly on top of particles of 
the first layer (cf. structural analysis in previous work \cite{haertel2015}). This explains our results 
of the van Hove functions, where the direction perpendicular to the wall is the preferred direction for 
a small test particle to move. 

In order to obtain the forces that a particle has to overcome in order to move in a certain direction, 
we average all force components acting along the opposite direction over the hemisphere of the test 
particle that points in the corresponding direction. To be specific, for a direction given by a vector 
\begin{equation}
\vec a(\vartheta,\varphi)=\text{sin}(\vartheta)\text{cos}(\varphi)\vec e_\text x 
+ \text{sin}(\vartheta)\text{sin}(\varphi)\vec e_\text y + \text{cos}(\vartheta)\vec e_\text z 
\end{equation}
we calculate the average 
\begin{equation}
\langle f_{1,\circ}(z,\vartheta)\rangle 
= \langle f_{1,\perp}(z,\vartheta')\vec a(\vartheta',\varphi')\cdot\vec a(\vartheta,0) 
\rangle_{\vartheta',\varphi'} , 
\end{equation}
where $f_{1,\perp}(z,\vartheta')$ is the normal force on the surface along direction $\vartheta'$ 
as plotted in Fig.~\ref{fig:ForceDistributionAndCaps}(a). The average 
$\langle\cdot \rangle_{\vartheta',\varphi'}$ is taken over the angles $\vartheta'$ and $\varphi'$ 
denoting a hemisphere around the direction given by $\vartheta$. Examples of such hemispheres are 
shown in Fig.~\ref{fig:ForceDistributionAndCaps}(c) for $\vartheta=0$ and $\vartheta=\pi/4$. 
We call $\langle f_{1,\circ}(z,\vartheta)\rangle$ the cross section force. 

Results for the cross section forces are shown in Fig.~\ref{fig:ForceDistributionAndCaps}(b) as 
functions of the direction for different particle positions $z$. The curves represent forces that a 
particle has to countervail in order to move into a certain direction. 
One can nicely see the change of the forces with the angle and the particle detachment. When a 
particle is close to contact (red line in Fig.~\ref{fig:ForceDistributionAndCaps}(b)), the cross 
section force exhibits a strong minimum for the direction pointing towards the wall ($\vartheta=\pi$). 
Note that the particle hardly can move into that direction due to the repulsion from the wall 
that is not included in the particle-particle force considerations. In the opposite direction 
($\vartheta=0$) a subtle local minimum occurs and there is a maximum into a tilted direction 
($\vartheta\approx \pi/4$) roughly denoting the direction towards the neighboring particles in the 
second layer (compare also Fig.~\ref{fig:intro}(b)). As the test particle is moved away from the wall 
towards the middle of the first two density layers, the minimum in the direction perpendicular away 
from the wall ($\vartheta=0$) becomes even more pronounced, whilst the maximum shifts towards the 
parallel direction (see, e.g., the cases represented by the yellow/green lines). 

Note that the cross section force cannot be calculated from a one-particle density, but at least a 
two-body description as presented in previous work \cite{haertel2015} is needed, because the anisotropy of 
the force distribution is crucial.

\subsection{Collision frequency in the hard-sphere limit}

\begin{figure}
\centering
\includegraphics[width=8.5cm]{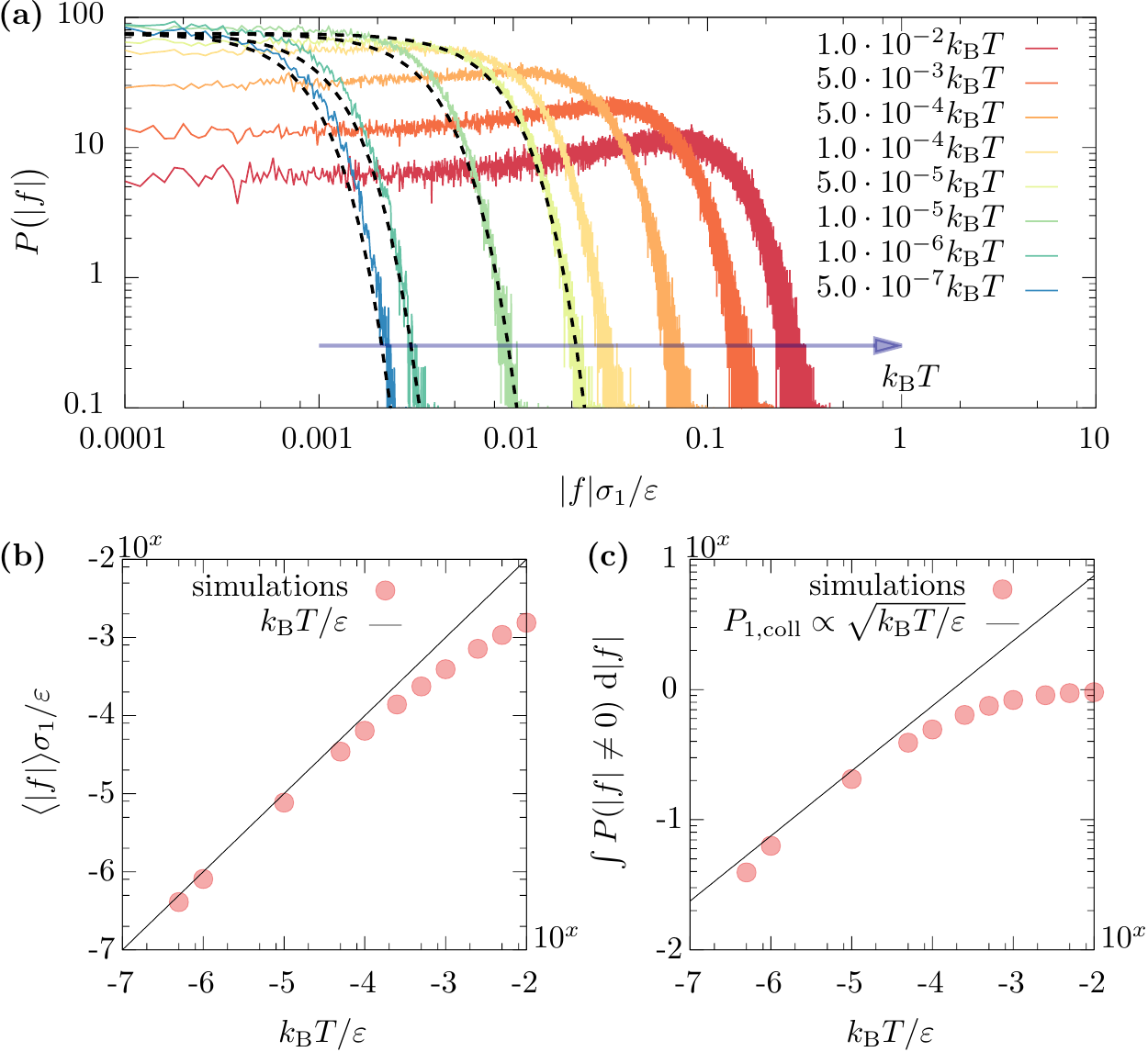}
\caption{\label{fig:ForceDistribution}(a) Force distributions $P(|f|)$ depending on the 
absolute value of interaction forces $|f|$ at a packing fraction of $\phi=0.54$ for different 
temperatures $T$. The colored lines are results of simulations and the black dashed lines are fits 
according to Eq.~(\ref{eq:BoltzmannFactorFit}). The factor according to 
Eq.~(\ref{eq:BoltzmannFactorFit}) has a constant value of $A=75$. (b) Average of the absolute value 
of the force as a function of temperature, determined from simulations. The black line is proportional 
to $k_\text B T/\varepsilon$. (c) Fraction of non-zero forces, i.e., the normalization of the 
distributions from (a), where the integral runs only over all non-zero forces. Points are 
from simulations and the line is proportional to $\sqrt{k_\text B T/\varepsilon}$.} 
\end{figure} 

Besides the above discussed directional anisotropies, another ingredient matters for the interaction 
between the particles. In a perfect hard-sphere system the particles only interact upon collision 
and are force free most of the time. In order to determine how often such collision events occur, 
we analyze the distribution of the absolute value of the interaction force for our soft-sphere system 
where the overlapping energy is well-defined and then approach the hard-sphere limit. 

The Boltzmann factor for two particles of the species $\nu$ and $\nu'$ with the interaction 
potential $u_{\nu\nu'}(d)$ is proportional to the probability of finding those two particles 
with an overlap $d$. It is given by 
\begin{equation}
P_{\nu\nu'}(d ) \propto \text{exp}\left[ -\frac{u_{\nu\nu'}(d)}{k_\text B T}\right]
= \text{exp}\left[ -\frac{|f|^2\sigma_{\nu\nu'}^2}{2 \varepsilon  k_\text B T}\right] ,
\label{eq:BoltzmannFactorOverlap}
\end{equation}
where $f=\frac{\epsilon}{\sigma_{\nu\nu'}}\left(1-\frac{\Delta}{\sigma_{\nu\nu'}}\right)$ 
is the force according to our harmonic model potential as given in Eq.~(\ref{eq:pair_potential}). 
Therefore, the last term in Eq.~(\ref{eq:BoltzmannFactorOverlap}) is the probability how often a 
certain force occurs.

In Fig.~\ref{fig:ForceDistribution}(a) we show the force distribution obtained from simulations at 
different temperatures. For low temperatures the simulation results can be fitted with 
a Boltzmann-factor for the mixture that we employ in simulations, i.e., 
\begin{equation}
P(|f|) = \frac{A\sigma_1}{4 \varepsilon} \sum\limits_{\nu,\nu'=1,2} 
\text{exp}\left[ -\frac{|f|^2\sigma_{\nu\nu'}^2}{2 \varepsilon  k_\text B T}\right] ,
\label{eq:BoltzmannFactorFit}
\end{equation}
where $A$ is a dimensionless fitting parameter. For temperatures below $10^{-4}k_\text B T$, 
the fits in Fig.~\ref{fig:ForceDistribution}(a) become sufficiently good and the factor of the 
distribution stays constant. Therefore, if we calculate the first moment of the distribution, 
$\langle |f| \rangle \equiv \langle |f|,P(|f|) \rangle$, the result 
(Fig.~\ref{fig:ForceDistribution}(b)) is inversely proportional to the temperature, as it should be. 
Since the force distributions follow a Gaussian, the expectancy value of a modified normalization only 
over non-zero forces must be proportional to $A\sqrt{k_\text BT}$. 
The simulation results are shown in Fig.~\ref{fig:ForceDistribution}(c) together with a square root 
power-law, $P_\text{1,coll}\propto\sqrt{k_\text B T}$, which can be applied as a prediction of the 
fraction of interacting (or colliding) particles at low temperatures, where sufficiently below 
jamming particles exhibit only one or zero overlaps. 

If the factor $A$ is known, one can then deduce the collision probability 
\begin{equation}
 P_{1,\text{coll}} = \frac{A}{8} 
 \sqrt{\frac{{2\pi k_\text B T}}{\varepsilon}}
 \sum\limits_{\nu,\nu'=1,2}\frac{\sigma_{1}}{\sigma_{\nu\nu'}} . 
\end{equation}
This ingredient will be used in section \ref{sec:rw} to develop a random walk model for a single 
tracer particle in front of the wall, which interacts with its environment, just with a 
probability equal to $P_{1,\text{coll}}$. For our random walk models, 
we will determine the factor $A$ from fits to simulation data as shown in 
Fig.~\ref{fig:ForceDistribution}(a). Note that in principle $A$ for the hard sphere limit also 
could be estimated by calculating the average force expected from the bulk equation of state of 
hard spheres, i.e., obtained from DFT with the White Bear mark II functional \cite{hansen-goos_jpcm18_2006}.

\subsection{Memory effects}

\begin{figure*}
\centering
\includegraphics[width=16.0cm]{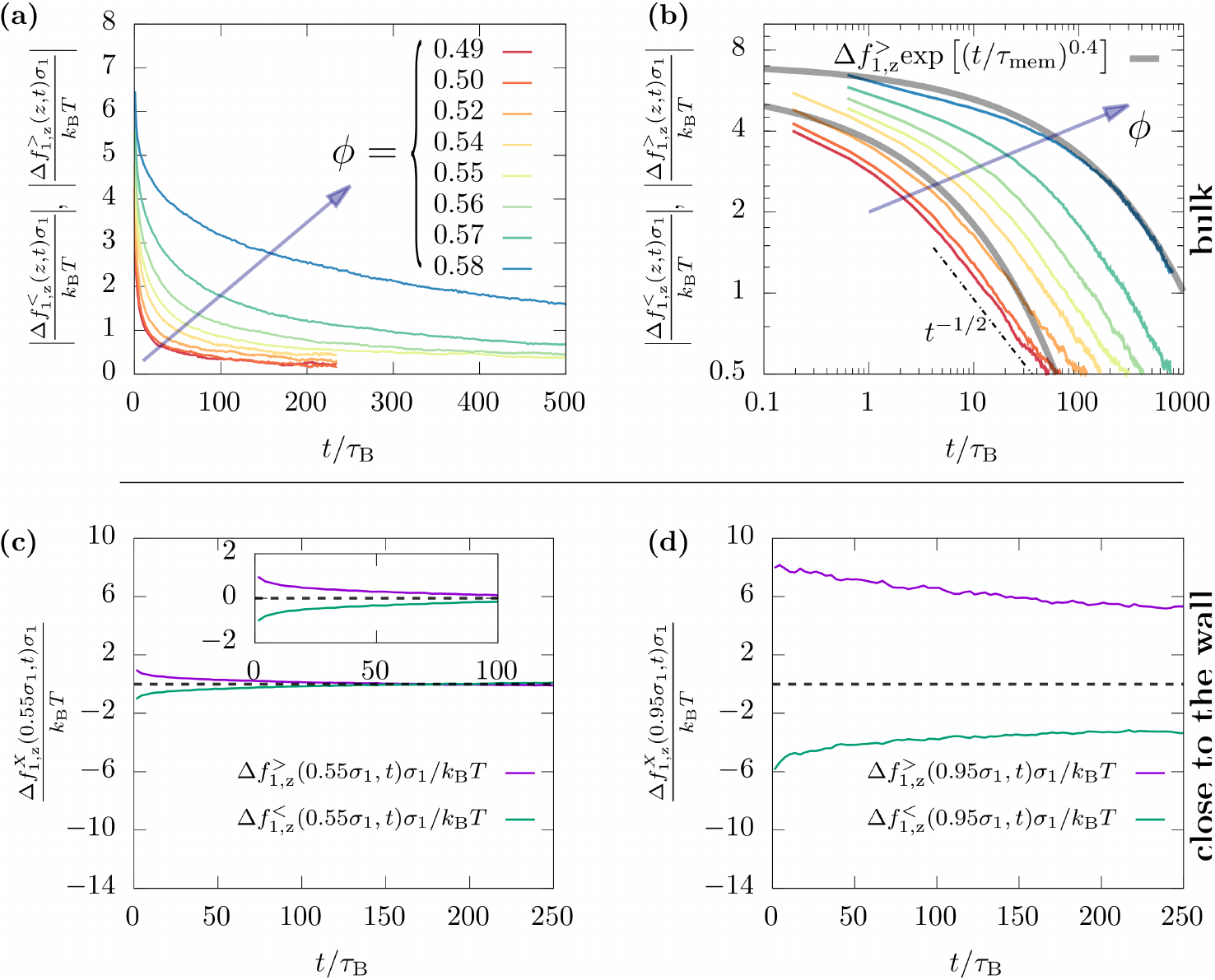}
\caption{\label{fig:HistoryDependencePlots} (a,b) Bulk force memories 
$\Delta f_{1,\text z}^<( z,t)$ and $\Delta f_{1,\text z}^>( z,t)$ defined as deviations of 
the forces from their average value for particles that have previously been closer to the wall 
or further away from the wall, respectively. Different packing fractions are considered. 
Since the bulk limit is considered, $z$ does not matter and the absolute value of the 
different force memories ($<$ or $>$) 
coincide. 
In (a) a linear and in (b) a double-logarithmic representation is shown, respectively. 
The black dashed line (b) is a power-law with exponent $-1/2$ and the thick gray lines are fits 
according to $\Delta f_{1,\text z}^> \text{exp}\left[(t/\tau_\text{mem})^{0.4}\right]$, 
with $\tau_\text{mem}\approx 187.5\tau_\text B$ (for $\phi=0.58$) or 
$\tau_\text{mem}\approx 6.3\tau_\text B$ (for $\phi=0.52$). (c,d) Force memory close to the wall 
in the case of the largest considered packing fraction ($\phi=0.58$) for small particles at 
(c) $z=0.55\sigma_1$ and (d) $z=0.95\sigma_1$. The dashed black lines indicate zero. }
\end{figure*} 

On the way to deduce atomistic descriptions of potential barriers that a particle needs to 
overcome in order to leave its cage into a given direction, we found out that for a small particle 
the escape from the cage along the $z$-axis is a probable process. As it will turn out, it is 
not sufficient to simply consider a mean potential force as an external field in order to describe 
the confinement by a local cage. The reason is that within such a simple mean field description 
there is no memory, i.e., no dependence on the history. Thus, we analyze the importance 
of memory effects in the following. 

In order to access the dependence on the history of a particle, we separately determine the forces 
for the following two groups of particles: First, particles that at a previous time $-t$ were at a 
position $z(-t)>z(0)$, i.e., further away from the wall than the current position. Second, 
particles that were at a previous time $-t$ at a position $z(-t)<z(0)$ closer to the wall. Then, 
the average forces on a particle are 
\begin{align} 
\label{eq:history_dependent_force_above}
f^>_{1,\text z} (z,t) &= 
\frac{1}{N_1^{'}}\sum\limits_{i=1}^{N_1^{'}} f_{1,\text z} \big(z(0)\big)\theta\big(z(-t)-z(0)\big)  , \\
f^<_{1,\text z} (z,t) &= 
\frac{1}{N_1^{'}}\sum\limits_{i=1}^{N_1^{'}} f_{1,\text z} \big(z(0)\big)\theta\big(z(0)-z(-t)\big) , 
\label{eq:history_dependent_force_below}
\end{align} 
where Eq.~(\ref{eq:history_dependent_force_above}) denotes the forces for particles that in the 
past have been further away from the wall, while Eq.~(\ref{eq:history_dependent_force_below}) 
gives the forces for spheres that previously have been closer to the wall. Again, $N_1^{'}$ is 
the number of the considered particles only. 

Furthermore, we introduce the deviations from the average value $\langle f_{1,\text z}(z)\rangle$ 
and call those functions the force memories, i.e., 
\begin{align} 
\Delta f^X_{1,\text z} (z,t) = f^X_{1,\text z} (z,t) - \langle f_{1,\text z} (z) \rangle  
\label{eq:history_dependent_force_delta} , 
\end{align} 
with the placeholder $X\in\{<,>\}$. In Fig.~\ref{fig:HistoryDependencePlots}(a) and (b) we plot 
the history-dependent force memories for bulk systems with different packing fractions in linear 
and double logarithmic representation, respectively. As the packing fraction is increased, two 
major characteristics determine the curves. The first is the starting point of the memory curve, 
i.e., $ \Delta f_{1,\text z}^>(z,t\to 0)$, which sets the magnitude of the force memory. The 
second is the timescale, on which the memory approximately decays. 
For the bulk, we pick out two systems, one being at intermediate packing fraction ($\phi=0.52$) 
and the other one at a large density ($\phi=0.58$) and fit the force memories via stretched 
exponential functions 
\begin{equation} 
\Delta f^>_{1,\text z}(z,t) 
= \Delta f_{1,\text z}^>(z) \text{exp}\left[ -(t/\tau_\text{mem})^\alpha \right] , \label{eq:stretchedExp}
\end{equation}   
where we fixed the exponent $\alpha=0.4$, in order to obtain reasonably comparable fits. 
For the two thick gray lines in Fig.~\ref{fig:HistoryDependencePlots}(b) the characteristic 
memory times are $\tau_\text{mem}=187.5\tau_\text B$ (for $\phi=0.58$) and 
$\tau_\text{mem}=6.3\tau_\text B$ (for $\phi=0.52$). 

Whilst the simulation curves seem to follow such a stretched exponential for short to intermediate 
times, for very old histories the memories behave like a power-law with an exponent $-1/2$. 
This is expected, since in an overdamped system our definition of the instantaneous force deviations 
are supposed to behave similar to the velocity-autocorrelation function 
(see, e.g., Ref.~\cite{Alder1970,Pomeau1975}) that in three dimensions is expected to decay with an 
exponent of $-3/2$ \cite{Ernst1984}. Since we restrict the diffusion direction to one direction, 
our curves should be comparable with the velocity-autocorrelation functions in one dimension 
with a decay exponent of $-1/2$ \cite{Pomeau1975}. 

The meaning of the force memory can be interpreted in the following way: as a particle leaves 
its initial location, in the case of dense systems it has to overcome an effective energy barrier 
that is caused by its cage. During a short-distance movement the particle already deforms its 
cage in such a way that the probability of finding a neighboring particle in the direction of 
its movement increases. At the same time, the probability to lack a neighbor in the opposite 
direction, i.e., behind the particle, increases. This causes the particle to feel an average 
force, the force memory, which acts against the direction of its motion. The memory effect is 
more pronounced for denser systems than for dilute suspension. The timescale of the memory 
depends on the timescale for relaxation processes in the corresponding systems. 

\begin{figure}
\centering
\includegraphics[width=8.0cm]{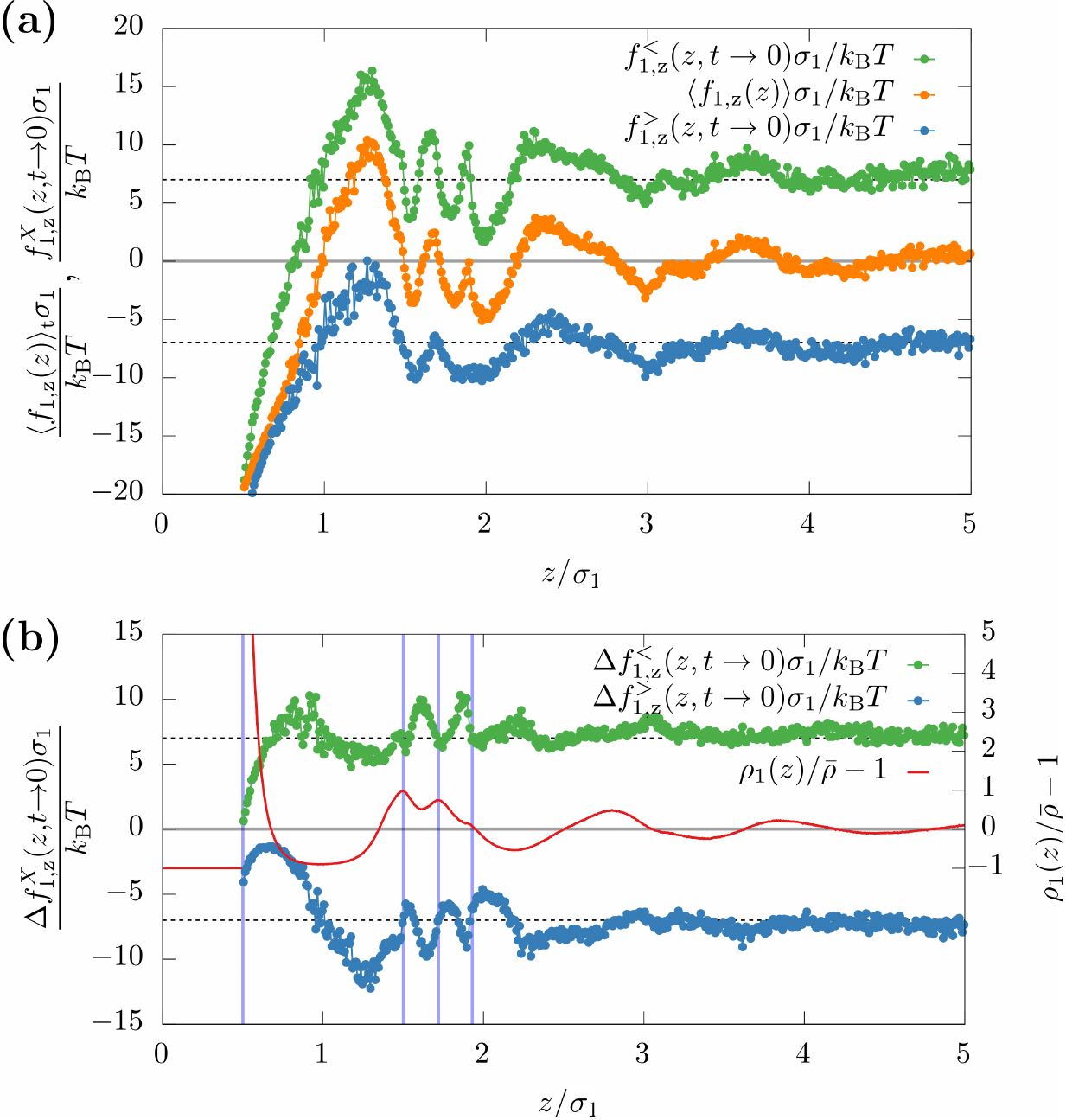}
\caption{\label{fig:ForceHistoryDependenceWithZ}(a) History-dependent forces for small 
particles at a packing fraction $\phi=0.58$ as a function of the distance $z$ above the 
wall for particles that previously have been closer (green line) or further away from 
the wall (blue line). The previous position was evaluated at a time $-t =- 1/3\tau_\text B$ 
which approximately corresponds to the limit $t\to 0$. The central orange line corresponds 
to the average force profile $\langle f_{1,\text z} (z) \rangle$. (b) Force memories, i.e., 
deviations from the average force, of the same data as in (a). In addition, the local 
density profile $\rho_1(z)$ of the smaller particle species (red line, right axis) is 
shown. The vertical shallow blue lines mark local density maxima (or enhancements) 
in $\rho_1(z)$ and correspond to local drops of the absolute values of $\Delta f_{1,\text z}^X(z)$.
}
\end{figure}  

Whereas the force memory does not depend on the position of a bulk particle on average, it does close to the 
wall. Thus, in Fig.~\ref{fig:HistoryDependencePlots}(c,d) we show $\Delta f^X_{1,\text z}(z,t)$ 
for a packing fraction $\phi=0.58$ at wall contact and in between the first two layers. One can 
nicely see the quantitative difference between the memories at the two investigated positions. 
For a particle, which is in the first layer and therefore very close to the wall, the force 
memory has only very subtle impact (Fig.~\ref{fig:HistoryDependencePlots}(c)). Contrarily, 
the memory for a particle in between two layers seems to be crucial 
(Fig.~\ref{fig:HistoryDependencePlots}(d)). 

In Fig.~\ref{fig:ForceHistoryDependenceWithZ}(a) the history-dependent forces at short times, 
i.e. $f^X_{1,\text z} (z,t \to 0)$, and for comparison the average force as functions of the 
position $z$ are shown. In Fig.~\ref{fig:ForceHistoryDependenceWithZ}(b) the respective force 
memories $\Delta f^X_{1,\text z} (z, t \to 0)$ are plotted. Here, the $t\to 0$ limit was 
determined by considering the corresponding functions at a time $t=\tau_\text B/3$ which is much 
shorter than the rearrangement time in all systems. 

Intriguingly, as marked by the vertical lines in Fig.~\ref{fig:ForceHistoryDependenceWithZ}(b) a 
connection between maxima of the local density (red line) and the magnitude of the force history 
(green and blue lines) becomes visible. Whenever a local density peak emerges, the magnitude of the 
memory is suppressed. The opposite is true for local density minima, where the magnitude of the 
corresponding force memories is enhanced. This shows that when particles reside inside a local 
density layer, they are supposed to be more stable at this position for longer times and 
individually occurring forces are similar. In return, this precipitates in the average of the 
force memory and therefore results in a less pronounced deviation from the total average. On 
the other hand, when a particle is on its way from one density maximum to another, it 
necessarily crosses a local minimum. On this crossing, its history has major influence on the 
average forces. For example, when the particle is moving in positive $z$-direction, it possibly 
leaves void space behind it, whereas in front of it a barrier of neighboring particles is blocking 
its path. Therefore, it is very likely that at such a position the memory has a large effect on 
the consecutive motion.

\section{One-particle random-walk models}
\label{sec:rw}

\begin{figure*}
\centering
\includegraphics[width=16.0cm]{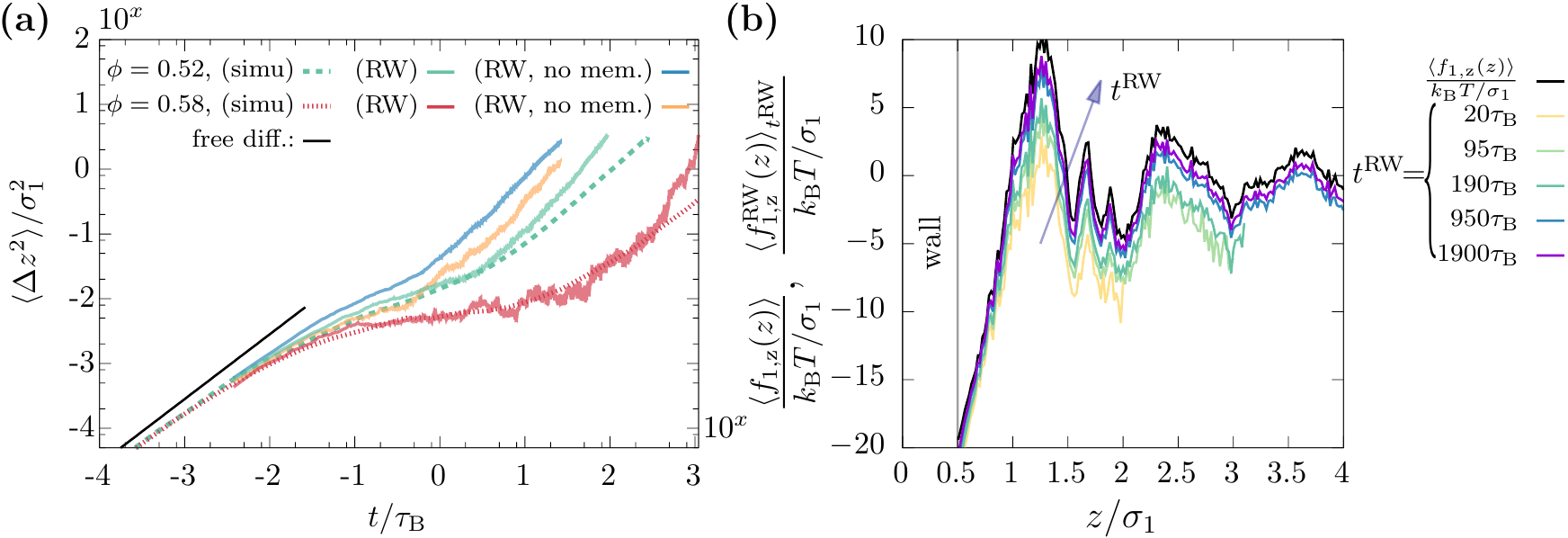}
\caption{\label{fig:MSDRWModel}(a) MSDs in $z$-direction for small particles starting at the wall. 
The plot shows data for two different packing fractions $\phi=0.52$ and $\phi=0.58$ in comparison 
to the data for a free particle (black solid line). The dashed lines are obtained from our 
multi-particle simulations, whereas the colored shallow lines are calculated from our random-walk 
models with (RW) and without (RW, no mem.) history-dependent memory. 
(b) Measured (average) position-dependent force in the random-walk model RW with incorporation of 
memory after different runtimes. In the limit of infinite runtime, the random-walk model RW predicts 
the same average force as our multi-particle simulation (black solid line). }
\end{figure*}

In this section, we test whether a random-walk model for a single particle can describe a cage 
breaking event.  We will not construct a model that is just based on an escape rate like Kramer's 
rate (see, e.g., \cite{haenggi1990}) 
or on an escape rate given in an free energy landscape as detmined in \cite{ekimoto_cpl577_2013}
but our random walks rely on the ingredients obtained in 
the previous section. 
To analyze the impact of the different ingredients, we construct 
random walk models where one particle in principle moves in three dimensions, 
though we use the models to describe the cage-breaking motion along the $z$-axis. 
We employ the diffusion coefficients as given by the temperature in order to determine 
the mean square of a step length and the average force profile in order to obtain a drift 
contribution in the case the particle is overlapping with another particle. The probability
for a collision as determined from simulation data is used as an additional input in order 
to decide whether there is any overlap at all. 
In the first model, no history-dependence is used, 
while in the second model we employ the memory function obtained from our simulations as an 
additional input. 
We will employ the test case, where a single small particle starts at the wall and moves 
towards the second layer in a perpendicular straight path.

\subsection{Random-walk model without memory} 

Our first random-walk model (RW, no mem.) does not incorporate any memory. It just employs the average 
force $\langle f_{1,\text z}(z)\rangle$ acting on a particle. The model consists of the following 
steps: 
\begin{enumerate}
\item Decide from an equally distributed random number and from the $z$-dependent collision 
	probability $P_{1,\text{coll}}(z)$, whether the particle collides with another particle. 
\item Calculate a three-dimensional diffusion step $\Delta \vec r^\text{step}$ within a 
	time $\Delta t^\text{step}$ via a diffusion coefficient according to free diffusion and an 
	external force in the case of a collision. 
\end{enumerate}
According to these steps, the external force vanishes if no collision occurs, but in the case 
of a collision an external force $\vec f^\text{step}=\vec f^\text{step,1}+\vec f^\text{step,2}$ 
with the following two contributions is applied: 
\begin{align}
\vec f^\text{step,1} &= 
-\frac{\Delta \vec r^\text{step}}{|\Delta \vec r^\text{step}|} 
\frac{\langle |\vec f_{1,\perp}(z)| \rangle}{P_\text{1,coll}(z)} , \\
\vec f^\text{step,2} &= 
\vec e_\text z \frac{\langle f_{1,\text z}(z)\rangle}{P_\text{1,coll}(z)} . 
\end{align}
Here, $\langle |\vec f_{1,\perp}(z)| \rangle$ is the average of the absolute (normal) forces 
acting on the particles surface at a position $z$. It is proportional to the local average 
pressure and can be calculated from the one-particle density together with the anisotropic pair 
correlations. The second force is the mean force, exerted by the structure of the neighboring 
particles. In both cases the normalization with the collision probability $P_\text{1,coll}(z)$ 
guarantees that the average force profile $\langle f_{1,\text{z}}(z) \rangle$ is recovered over time.

\subsection{Random-walk model with memory}

Our second random-walk model (RW) incorporates the memory of the history of the 
particle, namely the fact that it has started at the wall and, for this reason, is supposed 
to recognize the impact of the force memory $\Delta f^<_{1,\text z}(z,t)$. In this model the 
second force $\vec f^\text{step,2}$ of the model RW, no mem. is slightly modified such that it reads 
\begin{equation}
\vec f^\text{step,2} = \vec e_\text z \frac{  \langle f_{1,\text z}(z)\rangle 
+ \Delta f^<_{1,\text z}(z) \text{exp}\big[ -(t/\tau_\text{mem})^{0.4} \big] }{P_\text{1,coll}(z)} , 
\end{equation}
where we use the fits of the force memories from Eq.~(\ref{eq:stretchedExp}) as an input and 
the respective parameters as indicated in Fig.~\ref{fig:HistoryDependencePlots}(b).

\subsection{Comparison of the random-walk predictions and simulation results} 

In Fig.~\ref{fig:MSDRWModel}(a) we plot the MSD of small particles in $z$-direction for 
simulation data in comparison with results from our random-walk models for two different 
packing fractions. In case of the history-dependent model RW the respective memory times are 
$t_\text{mem}=187.5\tau_\text B$ for $\phi=0.58$ and $t_\text{mem}=6.3\tau_\text B$ for $\phi=0.52$. 
For the simple random-walk model RW, no mem. without memory, the curves deviate from the measured 
ones already at small runtime $t$. The deviation is less pronounced for the dilute system, where 
memory is supposed to play a smaller role than for systems close to the glass transition. The model 
RW (with memory) leads to reasonable predictions with, at least, a typical rearrangement 
for both packing fractions. 
The impact of memory in our model system is illustrated in Fig.~\ref{fig:MSDRWModel}(b), 
where we plot the average measured forces as a function of the particle position $z$ and of 
the runtime $\tau_\text{RW}$. For short runtime, the particle does not manage to hop over the 
first few particle layers, which results in truncated lines. However, for increasing runtime, 
the time-averaged forces from our random walk models converge, as expected, against 
$\langle f_{1,\text z}(z) \rangle$. This mimics the loss of memory.

\section{Conclusions}
\label{sec:conclusions}

In this work we explored the properties of a cage-breaking event in a bidisperse mixture of 
spheres in the vicinity of a wall. We studied the ingredients which are necessary to develop an 
effective one-particle random-walk description of the dynamics. By considering cages close 
to a wall and an escape route perpendicular to the wall, we guaranteed that errors that arise 
due to averaging quantities over differently oriented cages are small. 
We discovered that a random-walk model 
can describe the dynamics reasonably well if a suitable anisotropic force distribution, a 
collision frequency, and a memory function are used as input. While the force distribution and 
the collision frequency in principle can be obtained from theories on a level of two-particle 
structural correlations (as studied in previous works \cite{dietrich1996,haertel2015}), the 
memory of a trajectory depends on the motion of all particles in the surrounding cage. Besides 
the memory function, no other collective motion effects had to be included. By this, we 
demonstrated that the memory is essential to describe the dynamics in a dense system. 

In order to develop comparable random-walk models for rearrangements in bulk, one has to be 
careful with averages over different cage orientations. In principle, it should be possible to 
construct such random-walk models following three steps. First, all relevant (probable) cages in 
bulk must be identified (cf. \cite{Royall2015}). Second, the distributions of forces must be 
determined for all these cage configurations in such a way that the orientation is always fixed 
relative to one neighboring particle. Third, the average over orientations must be calculated. 
That way, the dynamics along the escape route and 
along other directions can be determined independently before taking an average instead of 
determining the dynamics after averaging over directions, which would correspond to the 
dynamics based on an isotropic mean field cage. Thus, errors can be avoided which result from 
the fact that the averaging over directions and the determination of escape dynamics obviously 
do not commute. 
It would be interesting to investigate how our approach 
compares to free energy landscapes of a cage \cite{ekimoto_cpl577_2013} or how 
our results can be 
built in or compared to the theories which employ isotropic cages, e.g., the theory where the 
cage breaking event is facilitated by the elasticity of the 
cage \cite{PhysRevE.92.052304} or mode coupling theory \cite{Gotze1992,Gotze2008}.

\section*{Acknowledgements}

M.K. and M.S. are supported within the Emmy Noether program of 
the Deutsche Forschungsgemeinschaft (Grant Schm 2657/2). 
A.H. acknowledges financial support from the DFG within the priority program SPP 1726 
(grant number SP 1382/3-1).


\providecommand{\newblock}{}

\end{document}